\documentstyle[aps,preprint,psfig]{revtex}

\begin{document}
\date{\today}
\draft
\title{The Effective Particle-Hole Interaction and the
       Optical Response of Simple Metal Clusters}

\author{W. Ekardt}

\address{
Fritz-Haber-Institut der Max-Planck-Gesellschaft,\\
Faradayweg 4-6, 14195 Berlin, Germany}

\author{J. M. Pacheco}

\address{
Departamento de Fisica da Universidade \\
3000 Coimbra, Portugal}

\maketitle

\begin{abstract}
Following
Sham and Rice [L. J. Sham, T. M. Rice, Phys. Rev. {\bf 144} (1966) 708]
the correlated motion of particle-hole pairs is studied, starting from
the general two-particle Greens function. In this way we derive a matrix
equation for eigenvalues and wave functions, respectively, of the general
type of collective excitation of a N-particle system. The interplay between
excitons and plasmons is fully described by this new set of equations.  As a
by-product  we obtain - at least {\it a-posteriori} - a justification
for the use of the TDLDA for simple-metal clusters.
\end{abstract}



\section{Introduction}

Though there is no exact theorem sanctioning the use of time-dependent
density functionals, the TDLDA (Time-Dependent Local Density Approximation)
as intuitively introduced by Zangwill and Soven in 1980\cite{z_s} turned out
to be working extremely well in the description of many-body effects
in atoms\cite{z_s}, molecules\cite{l_s}, clusters\cite{we1}, and
solids\cite{eg1}. A common  feature of
all these examples is the charge  density  character of these collective
states.
But besides this type of correlated particle-hole motion there is another one
well known from the optical properties of semiconductors, ionic solids,
and rare gases as  well, namely, the bound states of particles and holes,
 which generally occur if the attractive part of the general particle-hole
irreducible interaction\cite{noz} dominates. In metals this does not happen,
because the screened particle-hole interaction does not support bound states
for usual values of the Thomas-Fermi screening  vector\cite{mah}.
In small metal particles the situation is more subtle because  all states  are
size-quantized\cite{we2}; this implies that the effective particle-hole
interaction is anywhere between a semiconductor ($e^2/\epsilon r$, with
$\epsilon$ the static dielectric constant) and a metal
($e^2exp(kr)/r$, with $k$ the Thomas-Fermi screening vector).
For this reason, a more general theory is mandatory for the study of  the
optical absorption by collective states in metal clusters.
Recalling the
general discussion on the interplay between excitons and plasmons in the past
\cite{hor,egri}, the study of the general two-particle Greens function seems
to constitute a promising tool  to answer the question of whether or not the
oversimplifying TDLDA  has to be replaced by a more complete  theory ,
in order to obtain a satisfactory insight in the nature of collective states
in metal clusters. Starting from the method of Sham and Rice\cite{s_r}, a new
effective equation is derived in section two, for the calculation of the
optical spectra in metal clusters. Using the well-established  jellium
model\cite{we2},  these equations are solved for a number of
Na-jellium clusters in section three. Finally, the conclusions and  the
comparison with alternative models of calculating the optical response in
metal clustes are presented in section four.

\section{Theory}

We are interested in investigating the properties
of excitons and plasmons in general many-electron systems.
Therefore, our starting point is the full two-particle Green's function,
suitable to describe excitations of such a
system involving the correlated motion of two quasi-particles.
It will be shown that a unified description of excitonic and
plasmonic excitations leads to a matrix equation which is formally
similar to the well-known RPAE equations in atomic physics,
with the essential difference of a better description of exchange,
which becomes now properly screened. This formal similarity is,
of course, not accidental, and reflects the many possible ways one can
derive the equations of motion for these excitations\cite{egri}.
\par
Following Sham and Rice\cite{s_r} we start by defining,
field-theoretically, the exciton/plasmon wave-function,
\begin{eqnarray}
f_{\lambda}(\vec{r_1},\vec{r_2},\omega)
&=&
< N , 0 | {\psi}^{\dagger}(\vec{r_2}),{\psi}(\vec{r_1}) | N , \lambda > =
\\ \nonumber
&=&
\lim_{ \eta \rightarrow 0} \frac{1}{2 \pi i}
\int d{\omega}_1 e^{i \eta \omega}
G_{0;\lambda} (\vec{r_1},{\omega}_1;\vec{r_2},{\omega}-{\omega}_1)
\end{eqnarray}
in terms of the appropriate function $G_{0;\lambda}$, which occurs in
the Fourier transform of the general two-particle Green's function and
is defined as follows\cite{s_r},
\begin{eqnarray}
G_{0;\lambda} (\vec{r_1},{\omega}_1;\vec{r_2},{\omega}_2)
&=&
\sum_{\alpha}
\frac {
< N , 0 | {\psi}(\vec{r_1}) | N + 1 , \alpha >
< N + 1 , \alpha | {\psi}^{\dagger}(\vec{r_2}) | N , \lambda >
      }
      {
\omega_1 + i \delta - E^0_N - E^{\alpha}_{N+1}
      } \\ \nonumber
&-&
\sum_{\alpha}
\frac {
< N , 0 | {\psi}^{\dagger}(\vec{r_2}) | N - 1 , \alpha >
< N - 1 , \alpha | {\psi}(\vec{r_1}) | N , \lambda >
      }
      {
\omega_2 + i \delta + E^0_N - E^{\alpha}_{N-1}
      }
\end{eqnarray}
As usual, $|N , 0>$ and $|N , \lambda>$ are the exact ground-state and
plasmon/exciton excited state of the many electron system, while
the $\psi$ are the usual field-operators. $\lambda$ represents
a possible set of conserved quantum numbers. $G_{0;\lambda}$
satisfies the following
homogeneous Bethe-Salpeter equation\cite{s_r}
\begin{equation}
G_{0;\lambda}(1,2) =
G(1,1')G(2',2) I(1',2';3',4')G_{0;\lambda}(4',1')
\end{equation}
where $1=(\vec{r_1},\omega_1)$, etc. and integration over repeated
indexes is assumed. $I$ is the irreducible
particle-hole interaction, which contains the essential ingredients
of the physical processses we want to describe. For $I$, we
consider the ladder approximation and neglect retardation
effects, including the basic interactions described
by the diagrams in fig.1. Upward pointing arrows describe
electron states, downward pointing arrows describe hole
states, the dashed line represents the bare
particle-hole interaction and the double dashed line
represents the screened particle-hole interaction. As is
well-known\cite{egri,s_r}, since
$I$ is an irreducible interaction, only diagram b) must be
screened (and vertex corrected).
\par
To proceed, we now explicitly write, for $I$,
\begin{equation}
I(\vec{r_1},\vec{r_2},\vec{r_3},\vec{r_4}) =
V_{bare}(\vec{r_1}-\vec{r_2})
\delta(\vec{r_2}-\vec{r_3})
\delta(\vec{r_1}-\vec{r_4})
-
W(\vec{r_1},\vec{r_2})
\delta(\vec{r_1}-\vec{r_3})
\delta(\vec{r_2}-\vec{r_4}),
\end{equation}
where $V_{bare}$ is the bare
interaction and $W$ is the screened interaction\cite{gw1}.
Considering the usual expansion for $f_{\lambda}$
\begin{equation}
f_{\lambda} = X_{ph} \phi_{p} \phi_{h} +
Y_{ph} {\phi_{h}}^{*} {\phi_{p}}^{*} \; \; ,
\end{equation}
where $X_{ph}$ and $Y_{ph}$ are the forwards-going and backwards-going
electron-hole amplitudes, respectively, whereas $\phi$ are
quasi-particle
wave-functions, and
inserting this expansion, together with the expression for
$I$ in eq.(2), and its solution in eq.(1) we obtain, after a
final frequency integration, the following matrix-equation,
\begin{equation}
\left(
\begin{array}{cc}
A & B \\
-B^* & -A^*
\end{array}
\right)
\left(
\begin{array}{c}
X \\
Y
\end{array}
\right)
=
E
\left(
\begin{array}{c}
X \\
Y
\end{array}
\right)
\end{equation}
where the matrices $A$ and $B$ are defined as
\begin{eqnarray}
A_{ph,p'h'} &=&
(e_p - e_h) \delta_{p,p'} \delta_{h,h'} +
<ph'|I|p'h> \; , \\
B_{ph,p'h'} &=& <pp'|I|h'h>  \; .
\end{eqnarray}
In eqs.(7),(8), $e_p$($e_h$), are quasi-particle (quasi-hole) energies,
whereas the matrix-elements between quasi-particle states read
\begin{equation}
<ph|I|p'h'> = \int d\vec{r_1} \int d\vec{r_2} \int d\vec{r_3} \int d\vec{r_4}
{\phi_p}^*(\vec{r_1}) \phi_{h}^*(\vec{r_2})
I(\vec{r_1},\vec{r_2},\vec{r_3},\vec{r_4})
\phi_{p'}(\vec{r_3}) \phi_{h'}(\vec{r_4})
\end{equation}
Because these  equations are formally identical to the
RPAE equations, except for the residual particle-hole interaction,
which has now a screened exchange term, we call the present results
the RPA-SE equations. To our knowledge, this is the first time
these equations have been explicitly obtained.
Furthermore, we would like to point out that these equations
are useful for the case of static screening, in which
retardation effects are neglected\cite{nt1}.
However, the modifications to be introduced in the present formalism, in order
to incorporate dynamical screening effects
are formally straightforward. We
defer the study of dynamical screening effects to a future
publication.
\par
In order to implement the RPA-SE, we start from a
independent particle picture. Traditionally, the Hartree-Fock
independent electron picture is the starting point. However,
and since we are interested in studying the optical response
of small clusters of simple metals, for which the Hartree-Fock
approximation is known to provide a rather poor starting
point, we start from a Density Functional Theory (DFT)
scheme in the Local Density Approximation (LDA).
Therefore, we compute the one-electron wave-functions and
energies in LDA to DFT, and correspondingly, the one-particle
Green's functions.
For the screened interaction $W$, we compute it in Hedin's GW
approximation\cite{hed1,hed2}, following the prescription proposed by
Hybertsen and Louie\cite{gw2}. In this framework, instead of
working out the complete self-consistent solution of the
GW equations, one starts from the LDA approximation and
carries out one iteration of the GW self-consistent scheme.
This corresponds to write, for $W$,
\begin{equation}
W(\vec{r_1},\vec{r_2}) = \frac{e^2}{|\vec{r_1}-\vec{r_2}|} +
\int d \vec{r_1} d \vec{r_2}
\frac{e^2}{|\vec{r_1}-\vec{r_3}|}
\chi(\vec{r_3},\vec{r_4})
\frac{e^2}{|\vec{r_4}-\vec{r_2}|}
\; ,
\end{equation}
where, the density-density correlation function $\chi$ is computed at the
level of TDLDA. In general
$\chi$ is a complex and energy-dependent function. In the static
approximation, we take $E=0$ and $\chi$ becomes a real function.
\par
Within TDLDA, or equivalently, the matrix RPA-LDA (see below),
the general form for the interaction $I$ is replaced
by the so-called residual interaction\cite{we1,cy1}. It reads,
according to the previous definitions,
\begin{eqnarray}
K(\vec{r_1},\vec{r_2},\vec{r_3},\vec{r_4}) &=&
\delta(\vec{r_3}-\vec{r_2})
\delta(\vec{r_4}-\vec{r_1})
[ \, V_{bare}(\vec{r_1},\vec{r_2})
+
\frac{dV_{XC}}{dn}
\delta(\vec{r_1}-\vec{r_2}) \, ]
\\ \nonumber
&=&
\delta(\vec{r_3}-\vec{r_2})
\delta(\vec{r_4}-\vec{r_1}) [ \; \; \,
\frac{e^2}{|\vec{r_1}-\vec{r_2}|} \; \; \, +
\frac{dV_{XC}}{dn} \delta(\vec{r_1}-\vec{r_2}) \,] \; .
\end{eqnarray}

\section{Results and discussion}

In this section we apply the new theory to a number of magic jellium clusters
of Na, and compare them to other models, mainly to the old fashioned
TDLDA\cite{we1}. Fig.2 gives the photoabsorption cross-section $\sigma$
per delocalized electron for various models of calculation. As a reference the
dashed line gives for the results of the matrix-RPA-LDA.  This curve is
(of course) completely equivalent to the TDLDA. The only difference is that
the latter calculates the dynamical polarizabilty and from this quantity the
optical absorption, whereas the former calculates eigenvalues and
eigenvectors,
respectively, and from these quantities the optical absorption
as described in detail by Yannouleas and Broglia \cite{cy1}. That means,
each spectral line is Lorentzian broadened (here with the parameter
$\eta$ chosen to be 0.1 eV).
\par
Next we solve the new equations, that means the irreducible
particle-hole interaction is considered as given by eq.(4), meaning that the
attractive part of the interaction is properly screened, and the repulsive
part is calculated as usual. The result is given as a full line in fig.2.
The difference to the matrix-RPA-LDA is not very spectular  (in the specific
case of $Na_8$), because the exchange part of the interaction is less
attractive. A little blue shift results for the main line of the optical
response.

Besides this effect there is a little redistribution of oscillator strength
on the blue  side-bands of the response on the high energy side of the main
line. This can be made more visible by an alternative representation of the
results in fig.2a, which shows the line-spectrum
(without Lorentzian broadening). That this is a general result can be seen
in fig.3, which shows our results
for $Na_{20}$. In addition, the free-particle distribution of oscillator
strength is given by crosses: upon inclusion of the
electron-electron-interaction in the optical response the two strong lines
which have been termed in the past as "fragmented plasmon "
do not change very much if one includes screening in the exchange part of the
general interaction. As it was the case for $Na_8$ the main change occurs in
the high-energy wing of the spectrum.

As one can see in the next figures 4 and 5 for $Na_{58}$ and
$Na_{92}$ this is a general result: the proper screening of the exchange part
is necessary on general reasons, but the detailed numerical consequences are
rather negligible. In a sense this is a  gratifying result,
because it shows that still the simple TDLDA is the most efficient linear
response-model for a variety of cases (atoms, clusters, surfaces and bulk).

\section{Conclusions}

The main qualitative features of the optical response of metal clusters
are given by the overwhelming importance of the formation of the colllective
oscillation of all delocalized electrons: this phenomenon already present
in the
classical description of the optical properties \cite{we1} persists to exist
in a trully microscopic description \cite{we1} using the rather simple model
of the TDLDA; therefore it seems not too surprising that a true
many-body solution of the problem,
based on Feynman diagrams -- in the way proposed by Sham and Rice
\cite{s_r} -- does not change  very much our picture of the optical response.
But nevertheless a study along this line has to be done, because there is
still an
ongoing controversy about the right description of the effects of exchange in
this context \cite{madjet1}.

For this reason the marginal difference between TDLDA  and the BSE description
is highly welcome, because the latter is rather accurate but numerically
complex, whereas the former
is numerically easy to do but lacks a thorough justification. In this respect,
we have a similar situation as we have with the so-called
self-interaction correction SIC\cite{jmp1}, which has no sanctioning by any
exact theorem, but turns out to be numerically equivalent to a quasiparticle
calculation
at the level of the GW-approximation \cite{gw1} (which is considered to be
on a firmer theoretical basis!). The general ph-interaction seems to be
more important in clusters of other elements  (e.g. rare gases, ionic species,
or semiconductors), which will be studied next.

\begin{figure}
\ \vspace*{2cm}\\
Figure can be obtained from: \verb|ekardt@fhi-berlin.mpg.de|
\vspace*{2cm}\\
FIG.~1.
Graphical representation of I as used in this work. It originates from
all parts of the irreducible particle-hole interaction of first order: bare
Coulomb $+$ bare exchange after the exchange part has been properly screened.
According to the general rules for Feynman graphs part a) must NOT be
screened, because this would result in a reducible diagram (compare
ref.\cite{s_r}). Vertex corrections are not considered in this work.
\end{figure}

\newpage
\begin{figure}
\psfig{figure=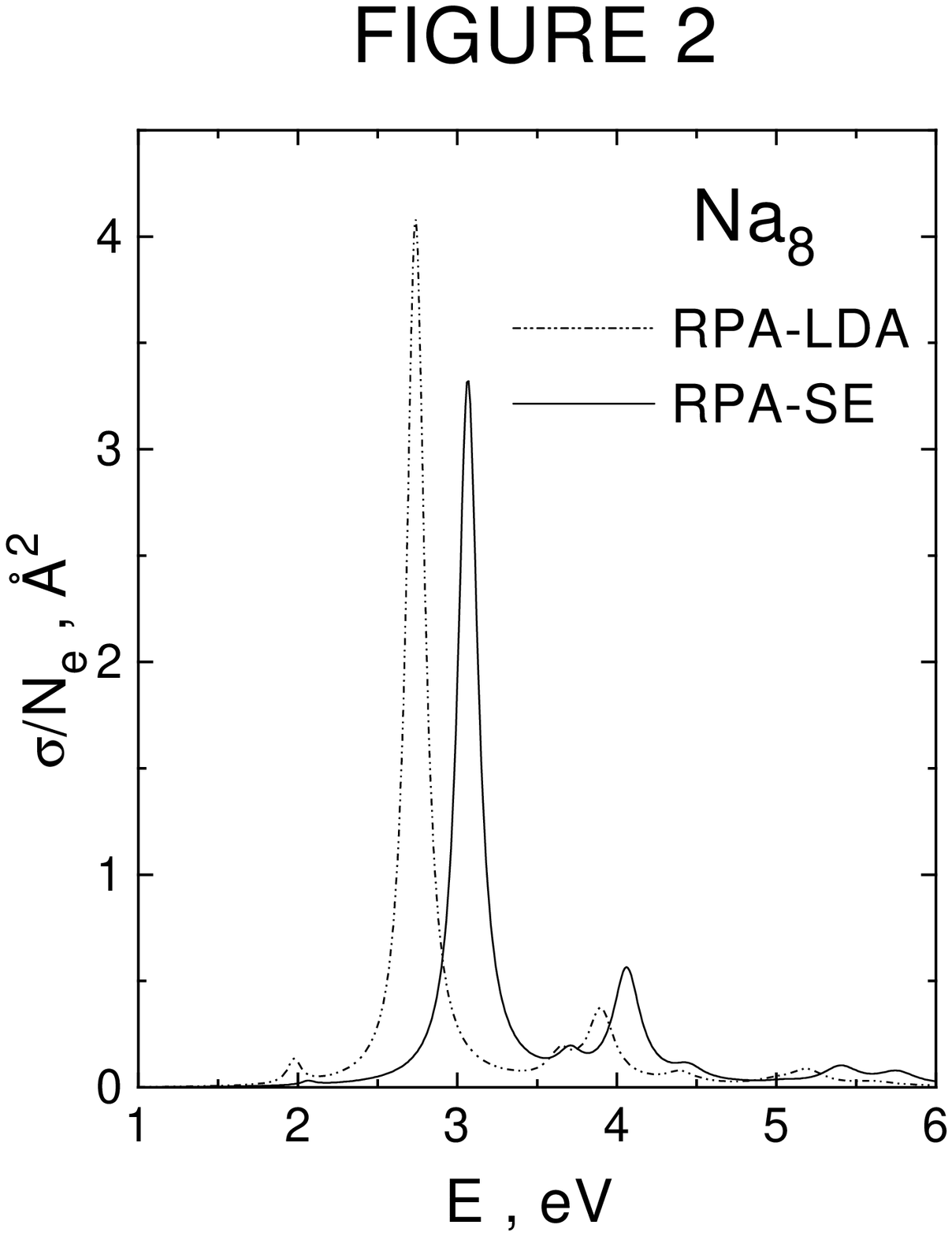,height=15cm}
FIG.~2.
Photoabsorption cross section per delocalized electron for jellium $Na_8$
in two different approximations: with a
dashed line, the matrix-RPA-LDA\cite{cy1}, which is equivalent to
TDLDA\cite{we1}; with a
continuous line, the RPA-SE as developed in this paper. The blue shift of the
main line results from the proper screening of the attractive part of the
general irreducible particle-hole interaction (\cite{s_r,gw1}).
For details, see main text.
\end{figure}

\newpage
\begin{figure}
\psfig{figure=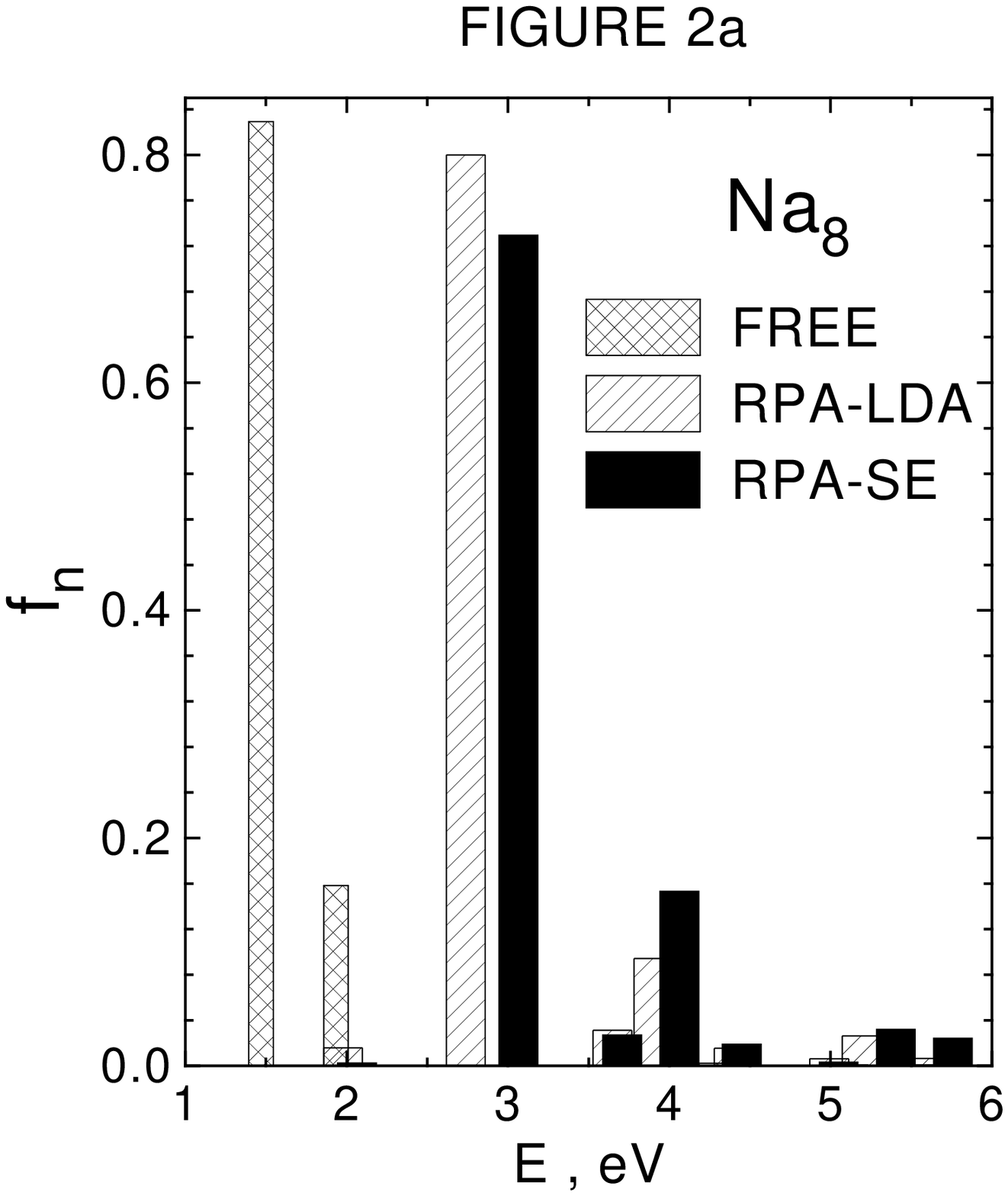,height=15cm}
FIG.~2a.
Frequencies and oscillator strength of various particle-hole transtions
in jellium $Na_8$
in three different approximations: crosses: the so-called free response
(meaning, that effects of the induced charge are neglected), the rest is the
same as in fig.1, but without Lorentzian broadening.  We chose this way of
representation, in order to make more visible the redistribution of
oscillator strength in the high-frequency part of the spectrum.
\end{figure}

\newpage
\begin{figure}
\psfig{figure=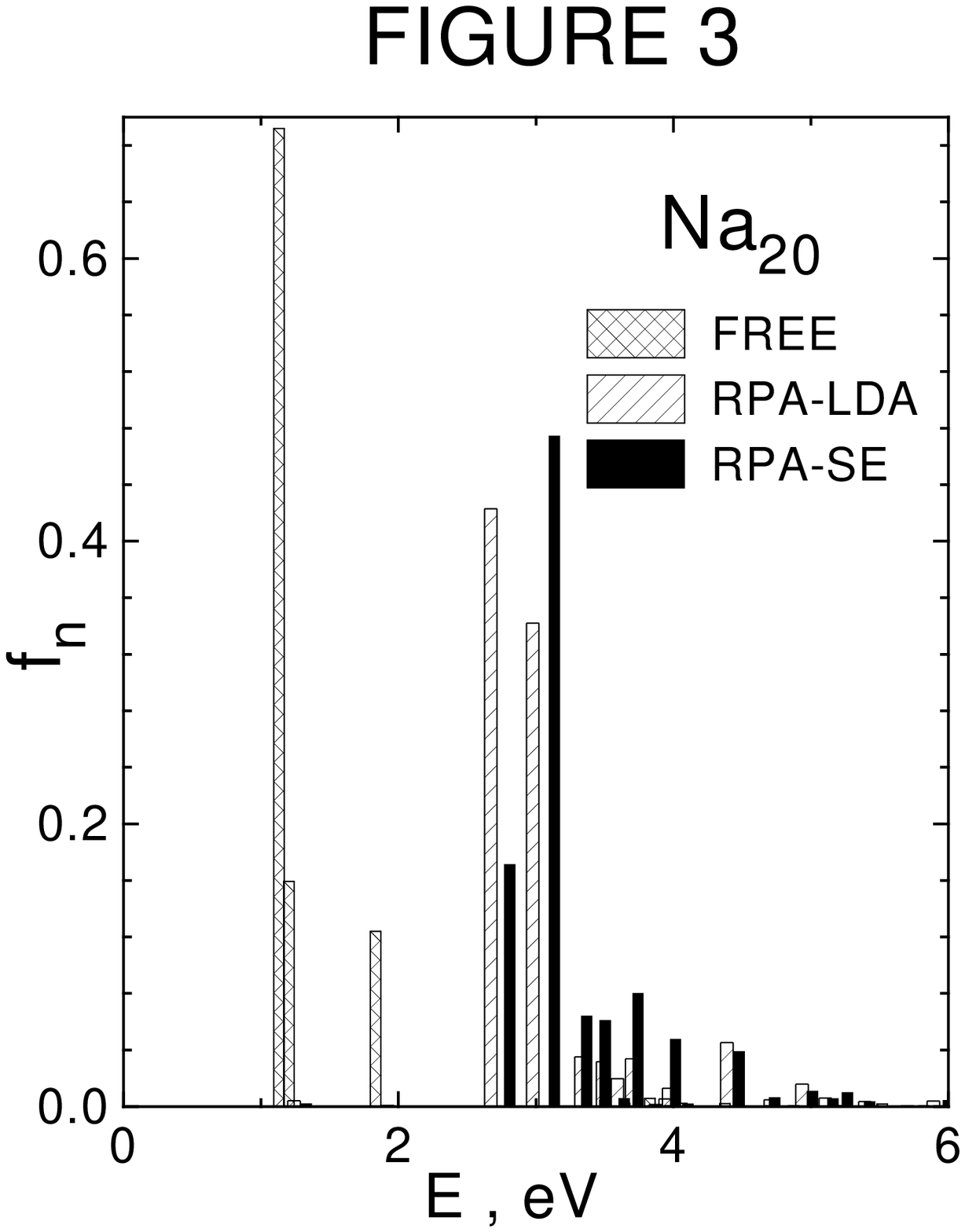,height=15cm}
FIG.~3.
The same as fig.2a, but for $Na_{20}$.
\end{figure}

\newpage
\begin{figure}
\psfig{figure=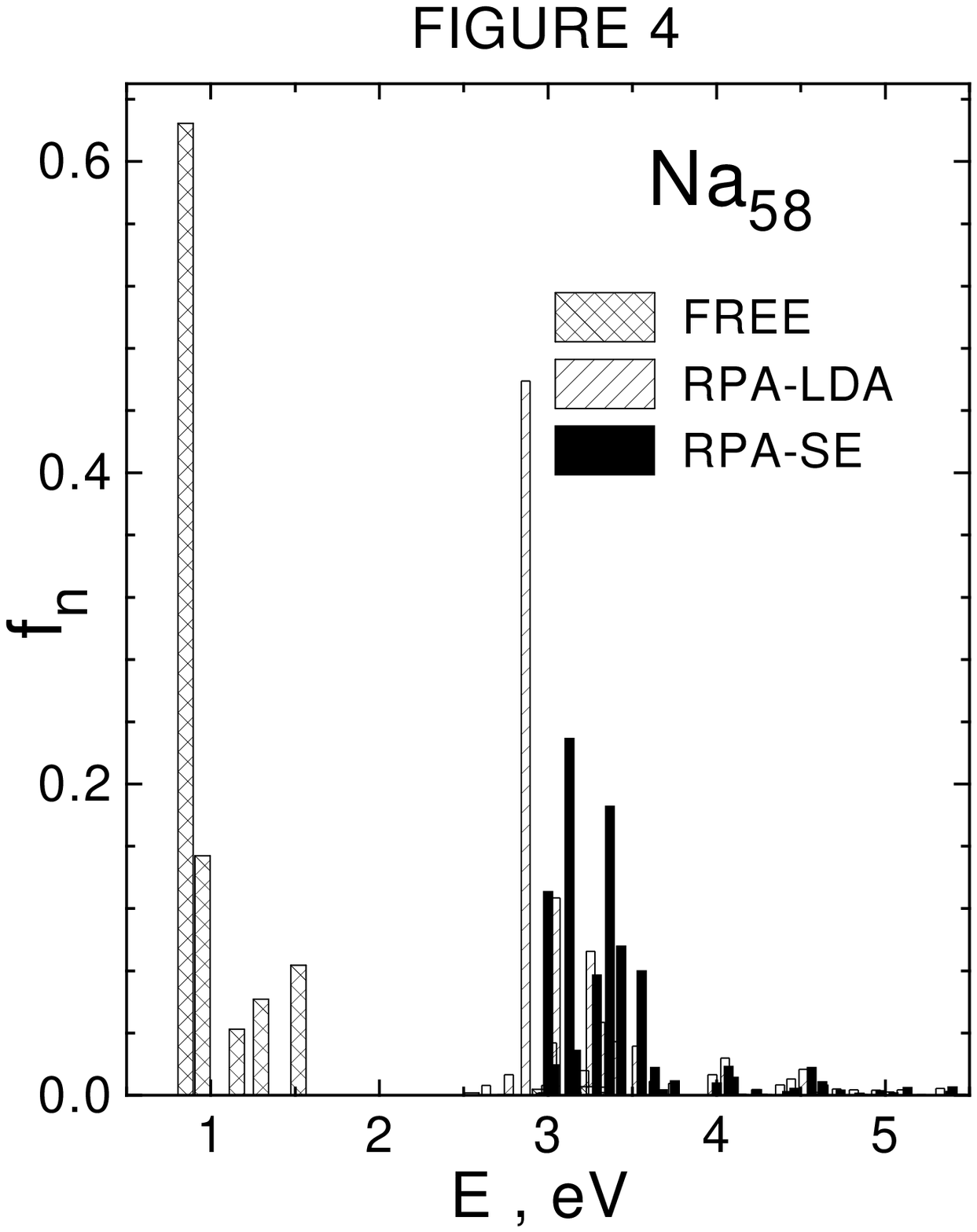,height=15cm}
FIG.~4.
The same as fig.2a, but for $Na_{58}$. This result is interesting, because
it shows clearly a fragmentation of the line shape in the collective
region around 3 eV. This has {\it not} been found in the preceding study of
C. Yannouleas, et al.\cite{cy2}.
\end{figure}

\newpage
\begin{figure}
\psfig{figure=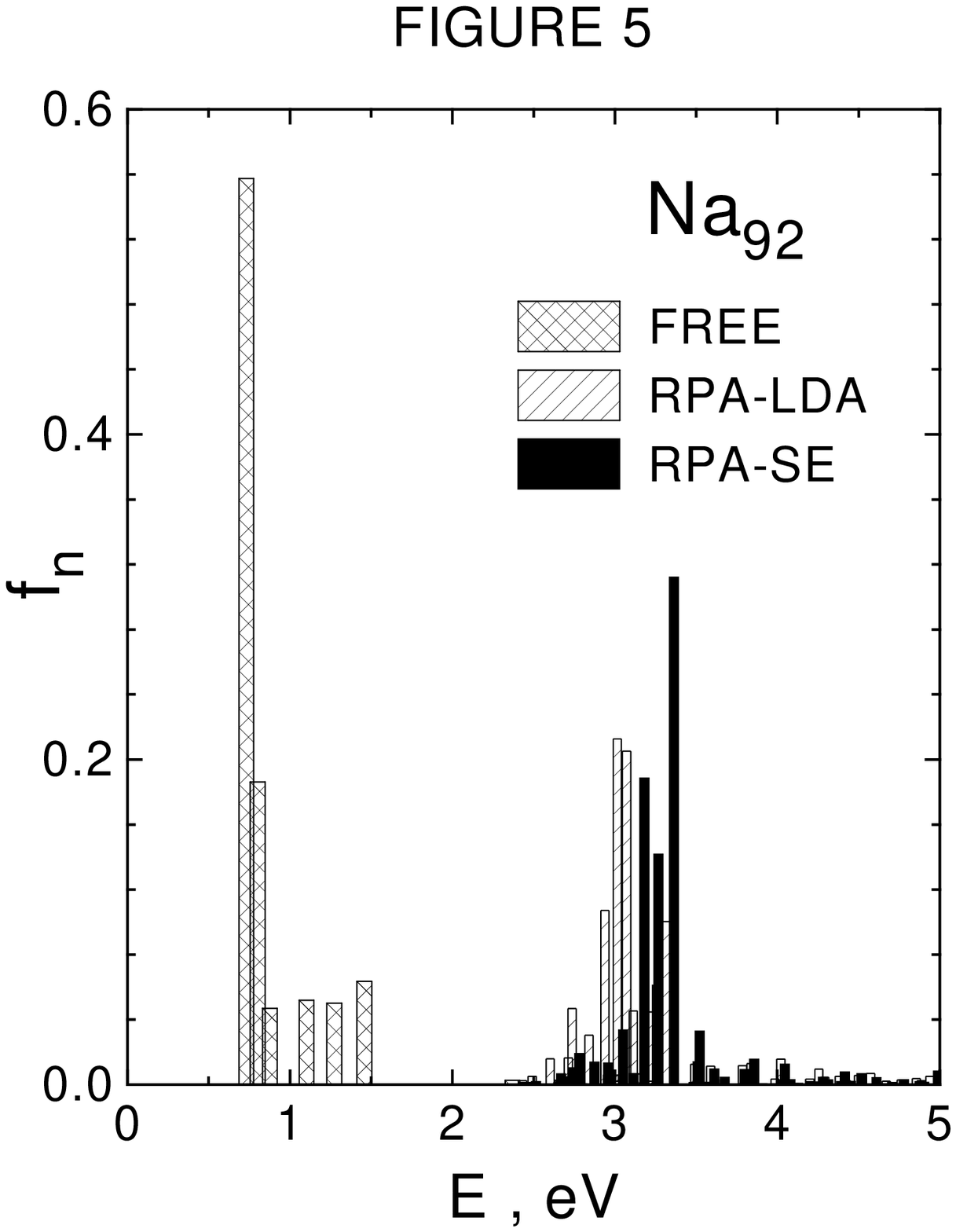,height=15cm}
FIG.~5.
The same as fig.2a, but for $Na_{92}$. Microscopically,
there {\it is} a fragmentation in the collective region. This, however,
should be hard to resolve experimentally (which agrees with
the interpretation of ref.\cite{cy2}).
\end{figure}


\begin{references}
\bibitem{nt1}     dynamical screening effects are known to be
                  important for a correct description of excitonic effects.
                  However, we will find that, in our case, the static
                  screening
                  provides a small contribution, and therefore we expect the
                  dynamical correction to be unimportant in the present
                  context.
\end{references}
\end{document}